\documentclass[prl,aps,showpacs,10pt,twocolumn]{revtex4-2}
\usepackage{amsmath}
\usepackage{epsfig}
\usepackage{times}
\def\onehalf{{\textstyle\frac{1}{2}}}

\def\vec#1{\mbox{\boldmath{$#1$}}}
\def\<{\left\langle}
\def\>{\right\rangle}
\begin{document}
\title{Exact invariants for a class of three-dimensional time-dependent
classical Hamiltonians}
\author{J\"urgen Struckmeier and Claus Riedel}
\affiliation{Gesellschaft f\"ur Schwerionenforschung (GSI),
Planckstr.~1, 64291~Darmstadt, Germany}
\date{Received 26 April 2000; revised manuscript received 8 June 2000}
\begin{abstract}
An exact invariant is derived for three-dimensional Hamiltonian systems
of $N$ particles confined within a general velocity-independent potential.
The invariant is found to contain a time-dependent function $f_{2}(t)$,
embodying a solution of a linear third-order differential equation whose
coefficients depend on the explicitly known trajectories of the
particle ensemble.
Our result is applied to a one-dimensional time-dependent
non-linear oscillator, and to a system of Coulomb interacting
particles in a time-dependent quadratic external potential.

\medskip\noindent
Published in: Phys.~Rev.~Lett.~\underline{85}, No.~18, 3830 (2000)
\end{abstract}
\maketitle
We consider a system of a non-relativistic ensemble of $N$ particles
of the same species moving in an explicitly time-dependent and
velocity-independent potential, whose Hamiltonian $H$ takes the form
\begin{equation}\label{ham0}
H=\sum_{i=1}^{N}\onehalf\Big[ p_{x,i}^{2}+p_{y,i}^{2}+
p_{z,i}^{2}\Big]+V(\vec{x},\vec{y},\vec{z},t)\,,
\end{equation}
with $\vec{x}$, $\vec{y}$, and $\vec{z}$ the $N$ component vectors
of the spatial coordinates of all particles.
It is hereby assumed that the system may be completely described
within $6N$-dimensional Cartesian phase-space spanned by the
$3N$ particle coordinates and their conjugate momenta.
From the canonical equations, we derive for each particle $i$
the equations of motion
\begin{equation}\label{speqm0}
\dot{x}_{i} = p_{x,i}\;,\qquad
\dot{p}_{x,i}= -\frac{\partial V(\vec{x},\vec{y},\vec{z},t)}{\partial x_{i}}\,,
\end{equation}
and likewise for the $y$ and $z$ degree of freedom.
The solution functions $\vec{x}(t)$, $\vec{y}(t)$, $\vec{z}(t)$,
and $\vec{p}_{x}(t)$, $\vec{p}_{y}(t)$, $\vec{p}_{z}(t)$ define a path
within the $6N$-dimensional phase-space that completely describes the
system's time evolution.
A quantity
\begin{equation}\label{invar}
I=I\left(\vec{x}(t),\vec{p}_{x}(t),\vec{y}(t),
\vec{p}_{y}(t),\vec{z}(t),\vec{p}_{z}(t),t\right)
\end{equation}
constitutes an invariant of the particle motion
if its total time derivative vanishes:
\begin{widetext}
\begin{align*}
\frac{d\mbox{} I}{d t}=\frac{\partial I}{\partial t}+\sum_{i=1}^{N}\bigg[
\frac{\partial I}{\partial x_{i}} \dot{x}_{i}+
\frac{\partial I}{\partial y_{i}} \dot{y}_{i}+
\frac{\partial I}{\partial z_{i}} \dot{z}_{i}+
\frac{\partial I}{\partial p_{x,i}}\dot{p}_{x,i}+
\frac{\partial I}{\partial p_{y,i}}\dot{p}_{y,i}+
\frac{\partial I}{\partial p_{z,i}}\dot{p}_{z,i}\bigg] = 0\,.
\end{align*}
We examine the existence of a conserved quantity (\ref{invar})
for a system described by (\ref{ham0}) with a special ansatz
for $I$ being at most quadratic in the momenta\cite{lewis-leach}
\begin{align}
I=\sum_{i}\Big[f_{2}(t)\left( p_{x,i}^{2}+p_{y,i}^{2}+
p_{z,i}^{2}\right)+f_{1}(x_{i},t)\,p_{x,i}
+g_{1}(y_{i},t)\,p_{y,i}+h_{1}(z_{i},t)\,p_{z,i}\Big]+
f_{0}(\vec{x},\vec{y},\vec{z},t)\,.\label{invar0}
\end{align}
The set of functions $f_{2}(t)$, $f_{1}(x_{i},t)$, $g_{1}(y_{i},t)$,
$h_{1}(z_{i},t)$, and $f_{0}(\vec{x},\vec{y},\vec{z},t)$ that render $I$
invariant are to be determined.
With the single particle equations of motion~(\ref{speqm0}),
a vanishing total time derivative of Eq.~(\ref{invar0}) means explicitly
\begin{align}
\sum_{i}\bigg[&\left(\dot{x}_{i}^{2}+\dot{y}_{i}^{2}+
\dot{z}_{i}^{2}\right)\frac{d\mbox{} f_{2}}{d t}+
\dot{x}_{i}\frac{\partial f_{1}}{\partial t}+
\dot{y}_{i}\frac{\partial g_{1}}{\partial t}+
\dot{z}_{i}\frac{\partial h_{1}}{\partial t}+
\dot{x}_{i}^{2}\frac{\partial f_{1}}{\partial x_{i}}+
\dot{y}_{i}^{2}\frac{\partial g_{1}}{\partial y_{i}}+
\dot{z}_{i}^{2}\frac{\partial h_{1}}{\partial z_{i}}+
\dot{x}_{i}\frac{\partial f_{0}}{\partial x_{i}}\notag\\
&+\dot{y}_{i}\frac{\partial f_{0}}{\partial y_{i}}
+\dot{z}_{i}\frac{\partial f_{0}}{\partial z_{i}}
-\left(2f_{2}\dot{x}_{i}+f_{1}\right)\frac{\partial V}{\partial x_{i}}-
\left(2f_{2}\dot{y}_{i}+g_{1}\right)\frac{\partial V}{\partial y_{i}}
-\left(2f_{2}\dot{z}_{i}+h_{1}\right)\frac{\partial V}{\partial z_{i}}\bigg]+
\frac{\partial f_{0}}{\partial t}=0\label{deritot}\,.
\end{align}
\end{widetext}
We may arrange the terms of this equation with regard to their power
in the velocities $\dot{x}_{i}$, $\dot{y}_{i}$, and $\dot{z}_{i}$.
Eq.~(\ref{deritot}) must hold independently of the specific
phase-space location of each individual particle $i$.
Therefore, the coefficients pertaining to the velocity
powers must vanish separately for each index $i$.
The condition for the terms proportional to $\dot{x}_{i}^{2}$ is
\begin{displaymath}
\frac{\partial f_{1}(x_{i},t)}{\partial x_{i}} +
\frac{d\mbox{} f_{2}(t)}{d t} = 0\,,
\end{displaymath}
and similarly for the functions $g_{1}$ and $h_{1}$.
It follows that $f_{1}(x_{i},t)$, $g_{1}(y_{i},t)$, and $h_{1}(z_{i},t)$
must be linear functions in $x_{i}$, $y_{i}$, and $z_{i}$, respectively
\begin{subequations}
\label{f1}
\begin{align}
f_{1}(x_{i},t) & = -\dot{f}_{2}(t)\,x_{i} + b_{x,i}(t)\tag{\ref{f1}a}\\
g_{1}(y_{i},t) & = -\dot{f}_{2}(t)\,y_{i} + b_{y,i}(t)\tag{\ref{f1}b}\\
h_{1}(z_{i},t) & = -\dot{f}_{2}(t)\,z_{i} + b_{z,i}(t)\tag{\ref{f1}c}\,,
\end{align}
\end{subequations}
with $b_{x,i}(t)$, $b_{y,i}(t)$, and $b_{z,i}(t)$ defined as arbitrary
functions of time only.

The terms of Eq.~(\ref{deritot}) that are linear in $\dot{x}_{i}$ sum up to
\begin{equation}\label{linear}
\frac{\partial f_{1}}{\partial t}+
\frac{\partial f_{0}}{\partial x_{i}} - 2f_{2}(t)\,
\frac{\partial V}{\partial x_{i}}=0\,.
\end{equation}
In order to eliminate $\partial f_{1} / \partial t$ from
(\ref{linear}), we calculate the partial time derivative
of $f_{1}(x_{i},t)$ from Eq.~(\ref{f1}a)
\begin{equation}\label{df1dt}
\frac{\partial f_{1}}{\partial t} =
-\ddot{f}_{2}(t)\,x_{i}+\dot{b}_{x,i}(t)\,.
\end{equation}
Again, similar expressions apply to $\partial g_{1}/ \partial t$
and $\partial h_{1}/ \partial t$.
Inserting (\ref{df1dt}) into (\ref{linear}), and solving for the terms
containing the partial derivatives of the yet unknown but arbitrary
ancillary function $f_{0}(\vec{x},\vec{y},\vec{z},t)$, one obtains
the following three differential equations for $f_{0}$:
\begin{subequations}
\label{tdf1}
\begin{align}
\frac{\partial f_{0}}{\partial x_{i}} & =
\ddot{f}_{2}(t)\,x_{i}-\dot{b}_{x,i}(t)
+2f_{2}(t)\,\frac{\partial V}{\partial x_{i}}
\tag{\ref{tdf1}a}\\
\frac{\partial f_{0}}{\partial y_{i}} & =
\ddot{f}_{2}(t)\,y_{i}-\dot{b}_{y,i}(t)
+2f_{2}(t)\,\frac{\partial V}{\partial y_{i}}
\tag{\ref{tdf1}b}\\
\frac{\partial f_{0}}{\partial z_{i}} & =
\ddot{f}_{2}(t)\,z_{i}-\dot{b}_{z,i}(t)
+2f_{2}(t)\,\frac{\partial V}{\partial z_{i}}
\tag{\ref{tdf1}c}\,.
\end{align}
\end{subequations}
A function $f_{0}(\vec{x},\vec{y},\vec{z},t)$ with partial derivatives
(\ref{tdf1}) is obviously given by
\begin{widetext}
\begin{align}
f_{0}(\vec{x},\vec{y},\vec{z},t) =
2f_{2}(t)\,V(\vec{x},\vec{y},\vec{z},t)+\sum_{i}\Big[
\onehalf\ddot{f}_{2}\big(x_{i}^{2}+y_{i}^{2}+z_{i}^{2}\big)-
\dot{b}_{x,i}\,x_{i}-\dot{b}_{y,i}\,y_{i}-\dot{b}_{z,i}\,z_{i}\Big]
\,.\label{f0}
\end{align}
The remaining terms of Eq.~(\ref{deritot}) do not depend
on the velocities $\dot{x}_{i}$, $\dot{y}_{i}$, and $\dot{z}_{i}$.
With (\ref{f1}), these terms impose the following condition
for $I$ to embody an invariant of the particle motion
\begin{align}
\sum_{i}\bigg[
\Big(\dot{f}_{2}\,x_{i}-b_{x,i}\Big)\frac{\partial V}{\partial x_{i}}+
\Big(\dot{f}_{2}\,y_{i}-b_{y,i}\Big)\frac{\partial V}{\partial y_{i}}+
\Big(\dot{f}_{2}\,z_{i}-b_{z,i}\Big)\frac{\partial V}{\partial z_{i}}\bigg]+
\frac{\partial f_{0}}{\partial t}=0\,.\label{rem}
\end{align}
In order to express Eq.~(\ref{rem}) in a closed form for $f_{2}(t)$,
one has to eliminate $\partial f_{0} / \partial t$.
To this end, we calculate the partial time derivative of
Eq.~(\ref{f0}), i.e., the time derivative at fixed
particle coordinates $x_{i}$, $y_{i}$, and $z_{i}$:
\begin{align}
\frac{\partial f_{0}(\vec{x},\vec{y},\vec{z},t)}{\partial t}
=2\dot{f}_{2}(t)\,V
+2f_{2}(t)\,\frac{\partial V}{\partial t}
+\sum_{i}\Big[\onehalf\dddot{f_{2}}\big(x_{i}^{2}
+y_{i}^{2}+z_{i}^{2}\big)-
\ddot{b}_{x,i}\,x_{i}-\ddot{b}_{y,i}\,y_{i}-
\ddot{b}_{z,i}\,z_{i}\Big]\,.\label{df0dt}
\end{align}
Inserting Eq.~(\ref{df0dt}) into Eq.~(\ref{rem}), we finally
get a linear third-order differential equation
for $f_{2}(t)$ and the $b_{x,y,z;i}(t)$ that only depends
on the spatial variables of the particle ensemble
\begin{align}
2\dot{f}_{2}(t)V+2f_{2}(t)
\frac{\partial V}{\partial t}+
\sum_{i}\bigg[&\onehalf\dddot{f_{2}}(t)\big(x_{i}^{2}+y_{i}^{2}+z_{i}^{2}\big)
+\dot{f}_{2}(t)\left(x_{i}\frac{\partial V}{\partial x_{i}}+
y_{i}\frac{\partial V}{\partial y_{i}}+
z_{i}\frac{\partial V}{\partial z_{i}}\right)\notag\\
&+b_{x,i}\,\ddot{x}_{i}-\ddot{b}_{x,i}\,x_{i}+
b_{y,i}\,\ddot{y}_{i}-\ddot{b}_{y,i}\,y_{i}+
b_{z,i}\,\ddot{z}_{i}-\ddot{b}_{z,i}\,z_{i}\bigg] = 0\label{dgl1}\,.
\end{align}
\end{widetext}
At this point, it is helpful to review our derivation made so far.
Speaking of an invariant $I$ of the particle motion means explicitly
to pinpoint a quantity (\ref{invar}) that is conserved along the
phase-space path representing the system's time evolution.
This path is defined as the subset of the $6N$-dimensional phase-space
on which the equations of motion (\ref{speqm0}) are fulfilled.
In order to work out the invariant $I$ of the particle motion,
the equations of motion (\ref{speqm0}) have been
inserted into the expression for $dI/dt=0$ in Eq.~(\ref{deritot}).
This means that the domain of (\ref{deritot}), and hence the physical
significance of the subsequent equations (\ref{dgl1}) and (\ref{invar1}),
is restricted to the actual phase-space path.
Along the phase-space path, all terms of Eq.~(\ref{dgl1})
that depend on the particle trajectories are in fact functions
of the parameter $t$ only.
Accordingly, the potential $V(\vec{x}(t),\vec{y}(t),\vec{z}(t),t)$
and its derivatives are time-dependent
coefficients of an ordinary differential equation for $f_{2}(t)$.
In contrast to Ref.~\cite{lewis-leach}, Eq.~(\ref{dgl1}) is not
conceived as a partial differential equation for $V$ in our context.

With $f_2(t)$ representing a solution of (\ref{dgl1}),
the invariant $I$ then follows from (\ref{invar0}), (\ref{f1}),
and (\ref{f0}) together with the Hamiltonian (\ref{ham0}) as
\begin{widetext}
\begin{align}
I=2f_{2}(t)\,H+\sum_{i}\Big[
&-\dot{f}_{2}(t)\big( x_{i}p_{x,i}+y_{i}p_{y,i}+z_{i}p_{z,i}\big)
+\onehalf\ddot{f}_{2}(t)\big(x_{i}^{2}+y_{i}^{2}+z_{i}^{2}\big)\notag\\
&+b_{x,i}\,p_{x,i}-\dot{b}_{x,i}\,x_{i}
+b_{y,i}\,p_{y,i}-\dot{b}_{y,i}\,y_{i}+
b_{z,i}\,p_{z,i}-\dot{b}_{z,i}\,z_{i}\Big]\label{invar1}\,.
\end{align}
The invariant (\ref{invar1}) is easily shown to embody
a time integral of Eq.~(\ref{dgl1}) by calculating the total
time derivative of (\ref{invar1}), and inserting the single
particle equations of motion (\ref{speqm0}).
Hence, Eq.~(\ref{invar1}) provides a time integral of Eq.~(\ref{dgl1})
if and only if the system's evolution is governed by
the equations of motion (\ref{speqm0}).

From their definition in Eqs.~(\ref{f1}),
$b_{x,i}(t)$, $b_{y,i}(t)$, and $b_{z,i}(t)$ are
arbitrary functions of time that do not depend on $f_{2}(t)$.
As a consequence, the sums over the terms containing the
respective functions in Eq.~(\ref{dgl1}) must vanish separately:
\begin{align}
\dot{f}_{2}(t)\bigg( 2V + \sum_{i=1}^{N}\bigg[
x_{i}\frac{\partial V}{\partial x_{i}}+
y_{i}\frac{\partial V}{\partial y_{i}}+
z_{i}\frac{\partial V}{\partial z_{i}}\bigg]\bigg)
+2f_{2}(t)\,\frac{\partial V}{\partial t} +
\dddot{f_{2}}(t)\sum_{i=1}^{N}\onehalf\big( x_{i}^{2}+
y_{i}^{2}+z_{i}^{2}\big) = 0\,,\label{dgl1a}\\
b_{x,i}\,\ddot{x}_{i}-\ddot{b}_{x,i}\,x_{i} = 0\,,\qquad
b_{y,i}\,\ddot{y}_{i}-\ddot{b}_{y,i}\,y_{i} = 0\,,\qquad
b_{z,i}\,\ddot{z}_{i}-\ddot{b}_{z,i}\,z_{i} = 0\,,\qquad
i=1,\ldots,N\,.\tag{\ref{dgl1a}'}
\end{align}
We thus obtain the following distinct invariants
\begin{align}
I_{f_{2}} = 2f_{2}(t)\,H-\dot{f}_{2}(t)\sum_{i=1}^{N}
\big( x_{i}p_{x,i}+y_{i}p_{y,i}+z_{i}p_{z,i}\big)
+\ddot{f}_{2}(t)\sum_{i=1}^{N}\onehalf\big( x_{i}^{2}+y_{i}^{2}+
z_{i}^{2}\big)\,,\label{invar1a}\\
I_{b_{x,i}} = b_{x,i}\,p_{x,i}-\dot{b}_{x,i}\,x_{i}\,,\qquad
I_{b_{y,i}} = b_{y,i}\,p_{y,i}-\dot{b}_{y,i}\,y_{i}\,,\qquad
I_{b_{z,i}} = b_{z,i}\,p_{z,i}-\dot{b}_{z,i}\,z_{i}\,,\qquad
i=1,\ldots,N\,.\tag{\ref{invar1a}'}
\end{align}
\end{widetext}
Regarding Eq.~(\ref{dgl1a}), one finds that for the particular case
$\partial V / \partial t\equiv 0$, hence for autonomous systems,
$f_{2}(t)=\mbox{const.}$ is always a solution of Eq.~(\ref{dgl1a}).
For this case, the invariant (\ref{invar1a}) reduces to
$I_{f_{2}}\propto H$, thus provides the system's total energy,
which is a known invariant for Hamiltonian systems with no
explicit time-dependence.
Nevertheless, Eq.~(\ref{dgl1a}) also allows for
solutions $f_{2}(t)\ne\mbox{const.}$ for these systems.
We thereby obtain other non-trivial invariants for
autonomous systems that exist in addition to the invariant
representing the energy conservation law.

Eq.~(\ref{dgl1a}) can be significantly simplified for potentials $V$
that may be expressed as a sum of homogeneous functions
\mbox{$V=\sum_{m}V_{m}$}.
By definition, $V_{m}$ is referred to as homogeneous if for every real
$\lambda > 0$ and all $\vec{x}$, $\vec{y}$, and $\vec{z}$ the condition
\begin{displaymath}
V_{m}(\lambda\vec{x},\lambda\vec{z},\lambda\vec{z},t)=
\lambda^{k_{m}}V_{m}(\vec{x},\vec{z},\vec{z},t)
\end{displaymath}
is satisfied, $k_{m}$ specifying the degree of homogeneity of $V_{m}$.
With $V$ a sum of homogeneous functions, Euler's relation
may be written as
\begin{equation}\label{euler}
\sum_{i=1}^{N}\left[ x_{i}\frac{\partial V}{\partial x_{i}}+
y_{i}\frac{\partial V}{\partial y_{i}}+
z_{i}\frac{\partial V}{\partial z_{i}}\right] =
\sum_{m}k_{m}V_{m}\,.
\end{equation}
Using (\ref{euler}), the differential equation (\ref{dgl1a}) for
$f_{2}(t)$ finally reads for homogeneous potential functions $V_{m}$
\begin{equation}
2f_{2}\frac{\partial V}{\partial t}+
\dot{f}_{2}\sum_{m}\big(k_{m}\!+2\big)V_{m}+
\dddot{f_{2}}\sum_{i=1}^{N}\!\onehalf
\big(x_{i}^{2}\!+\!y_{i}^{2}\!+\!z_{i}^{2}\big)=0\,.\label{dgl1b}
\end{equation}
As a simple example, we investigate the one-dimensional non-linear
Hamiltonian system of a time-dependent ``asymmetric spring'', defined by
\begin{equation}\label{ham3}
H = \onehalf p^{2}+\onehalf\omega^{2}(t)\,x^{2}+a(t)\,x^{3}\,.
\end{equation}
The related equation of motion follows as
\begin{equation}\label{speq3}
\ddot{x}+\omega^{2}(t)\,x+3a(t)\,x^{2}=0\,.
\end{equation}
The invariant $I_{f_{2}}$ is immediately found writing down the general
invariant (\ref{invar1a}) for one dimension and one particle
with the Hamiltonian $H$ given by (\ref{ham3})
\begin{equation}\label{invar4}
I_{f_{2}} = f_{2}\left(p^{2}+\omega^{2}x^{2}+2ax^{3}\right) -
\dot{f}_{2}xp + \onehalf\ddot{f}_{2}x^{2}\,.
\end{equation}
The function $f_{2}(t)$ for this particular case is given
as a solution of the linear third-order differential equation
\begin{equation}\label{dgl3a}
\dddot{f_{2}}+4\dot{f}_{2}\omega^{2}+4f_{2}\omega\dot{\omega}+
x(t)\left(4f_{2}\dot{a}+10\dot{f}_{2}a\right)=0\,,
\end{equation}
which follows from (\ref{dgl1a}) or, equivalently, from (\ref{dgl1b}).
Since the particle trajectory $x=x(t)$ is explicitly contained in
Eq.~(\ref{dgl3a}), it must be known prior to integrating Eq.~(\ref{dgl3a}).
The trajectory is obtained integrating the equation of motion (\ref{speq3}).

We may easily convince ourselves that $I_{f_{2}}$ is indeed a
conserved quantity.
Calculating the total time derivative of Eq.~(\ref{invar4}),
and inserting the equation of motion (\ref{speq3}), we
end up with Eq.~(\ref{dgl3a}), which is fulfilled by definition
of $f_{2}(t)$ for the given trajectory $x=x(t)$.

The third-order equation (\ref{dgl3a}) may be converted into
a coupled set of first- and second-order equations.
With the substitution \mbox{$\rho_{x}^{2}(t)\equiv f_{2}(t)$}, the
second-order equation writes
\begin{equation}\label{dgl3b}
\ddot{\rho}_{x}+\omega^{2}(t)\,\rho_{x}-\frac{g_{x}(t)}{\rho_{x}^{3}}=0\,.
\end{equation}
Eq.~(\ref{dgl3b}) is equivalent to (\ref{dgl3a}), provided that
the time derivative of the function $g_{x}(t)$, introduced in
(\ref{dgl3b}), is given by
\begin{equation}\label{dgl3c}
\dot{g}_{x}(t) = -x(t)\left( 2\dot{a}\rho_{x}^{4} +
10a\rho_{x}^{3}\dot{\rho}_{x}\right)\,.
\end{equation}
Expressing the invariant (\ref{invar4}) in terms of $\rho_{x}(t)$,
we get inserting the auxiliary equation (\ref{dgl3b})
\begin{equation}\label{invar3}
I_{\rho_{x}} = {\left( \rho_{x}\,p-\dot{\rho}_{x}\,x\right)}^{2}+
\frac{x^{2}}{\rho_{x}^{2}}g_{x}(t)+2a(t)\,\rho_{x}^{2}\,x^{3}\,.
\end{equation}
The invariant (\ref{invar3}) reduces to the well-known
Lewis invariant \cite{lewis,cousny} for the time-dependent harmonic
oscillator if $a(t)\equiv 0$, which means that $g_{x}(t)={\rm const}$.
For this particular case, Eq.~(\ref{dgl3c}), and hence Eq.~(\ref{dgl3b}),
no longer depends on the specific particle trajectory $x=x(t)$.
Consequently, the solution functions $\rho_{x}(t)$ and
$\dot{\rho}_{x}(t)$ apply to all trajectories that follow from
$\ddot{x}+\omega^{2}(t)\,x=0$.
With regard to Eq.~(\ref{dgl1a}), we conclude that a decoupling from
the equations of motion (\ref{speqm0}) may occur for linear systems only.

A more challenging example is constituted
by an ensemble of Coulomb interacting particles of
the same species moving in a time-dependent quadratic external potential,
as typically given in the co-moving frame for charged particle beams
that propagate through linear external focusing devices
\begin{align}
V(\vec{x},\vec{y},\vec{z},t)=\sum_{i}\Big[
\onehalf&\omega_{x}^{2}(t)\,x_{i}^{2}+
\onehalf\omega_{y}^{2}(t)\,y_{i}^{2}\notag\\
+\,\onehalf&\omega_{z}^{2}(t)\,z_{i}^{2}+
\onehalf\sum_{j\ne i}\frac{c_{1}}{r_{i j}}\Big]\,,\label{effpot}
\end{align}
with
$r_{i j}^{2}={(x_{i}-x_{j})}^{2}+{(y_{i}-y_{j})}^{2}+{(z_{i}-z_{j})}^{2}$
and \mbox{$c_{1}=q^{2}/4\pi\epsilon_{0}m$},
$q$ and $m$ denoting the particles' charge and mass, respectively.
The equations of motion that follow from (\ref{speqm0})
with (\ref{effpot}) are
\begin{equation}\label{speqm}
\ddot{x}_{i}+\omega_{x}^{2}(t)\,x_{i}-c_{1}\sum_{j\ne i}
\frac{x_{i}-x_{j}}{r_{i j}^{3}}=0\,,
\end{equation}
and likewise for the $y$ and $z$ degrees of freedom.
We note that the factor $1/2$ in front of the Coulomb interaction term
in (\ref{effpot}) disappears since each term occurs twice in the
symmetric form of the double sum.

Equation~(\ref{effpot}) may be split into a sum of two
homogeneous functions, namely the focusing potential part with the
degree of homogeneity $k_{1}=2$, and the Coulomb interaction part,
the latter with the degree $k_{2}=-1$.
Consequently, the third-order differential equation (\ref{dgl1b})
for $f_{2}(t)$ reads (after rearranging)
\begin{widetext}
\begin{align}
\sum_{i}\Big[\dot{f}_{2}\sum_{j\ne i}\frac{c_{1}}{r_{i j}} +
&x_{i}^{2}\left(\dddot{f_{2}}+4\dot{f}_{2}\omega_{x}^{2}+
4f_{2}\omega_{x}\dot{\omega}_{x}\right)
+y_{i}^{2}\left(\dddot{f_{2}}+4\dot{f}_{2}\omega_{y}^{2}+
4f_{2}\omega_{y}\dot{\omega}_{y}\right)
+z_{i}^{2}\left(\dddot{f_{2}}+4\dot{f}_{2}\omega_{z}^{2}+
4f_{2}\omega_{z}\dot{\omega}_{z}\right)\Big] = 0\label{dgl2a}\,.
\end{align}
\end{widetext}
Of course, the same result is obtained if the potential function
(\ref{effpot}) is directly inserted into Eq.~(\ref{dgl1a}).

The invariant $I_{f_{2}}$ for a system determined by the
Hamiltonian (\ref{ham0}) containing the potential
(\ref{effpot}) is given by (\ref{invar1a}), provided that
$f_{2}(t)$ is a solution of (\ref{dgl2a}).

\begin{figure}[hbt]
\begin{center}
\epsfig{file=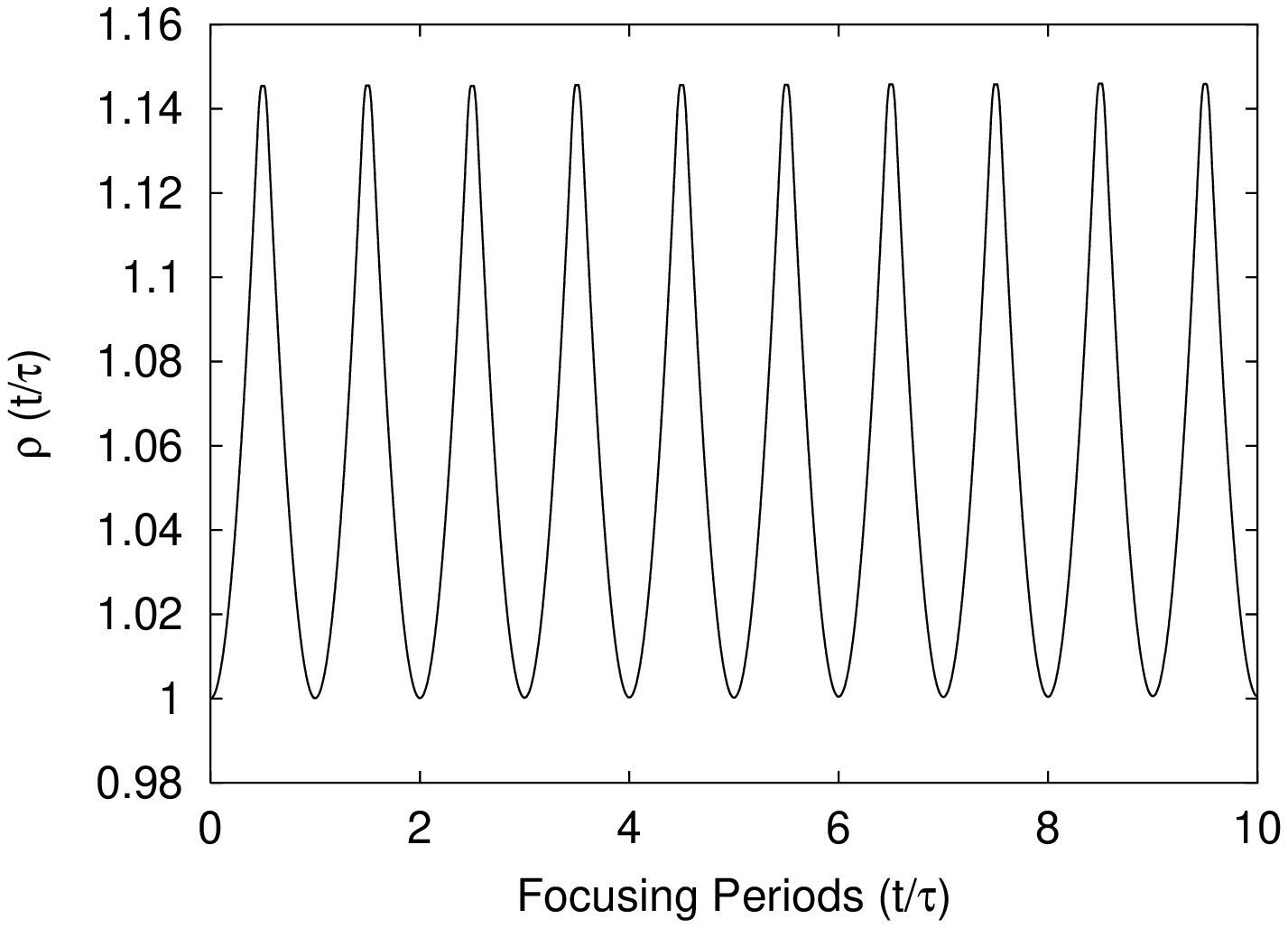,height=80mm,angle=-90}
\vspace{2mm}
\caption{$\rho$ versus time as obtained for 2500 simulation particles
in a 3D simulation of a periodic
focusing lattice with strong Coulomb interaction.
$\tau$ denotes the focusing period common to all three directions.}
\label{fig:f2quadr}
\end{center}
\end{figure}

Equation~(\ref{dgl2a}) may be cast into a compact form if the sums over the
particle coordinates are written in terms of ``second beam moments,''
denoted as $\< x^{2} \>$ for the $x$-direction.
The double sum over the Coulomb interaction terms constitutes
the electric field energy $W(t)$ of all particles
\begin{displaymath}
\< x^{2} \>\!(t) = \frac{1}{N}\sum_{i} x_{i}^{2}(t) \,,\qquad
W(t) = \frac{m}{2}\sum_{i}\sum_{j\ne i}\frac{c_{1}}{r_{i j}}\,.
\end{displaymath}
Substituting $\rho^{2}(t)\equiv f_{2}(t)$ and defining the
function $g=g(t)$ according to
\begin{align}
g(t)=&\<x^{2}\>\rho^{3}\left(\ddot{\rho}+\omega_{x}^{2}(t)\rho\right)+
\<y^{2}\>\rho^{3}\left(\ddot{\rho}+\omega_{y}^{2}(t)\rho\right)\notag\\
&+\<z^{2}\>\rho^{3}\left(\ddot{\rho}+\omega_{z}^{2}(t)\rho\right)
\,,\label{gvont}
\end{align}
the third-order equation (\ref{dgl2a}) can be transformed
into an equivalent coupled system of a second-order equation
for $\rho(t)$, and a first order equation for $g(t)$,
thereby eliminating the derivatives
$\dot{\omega}_{x,y,z}(t)$ of the lattice functions.
Solving (\ref{gvont}) for $\ddot{\rho}(t)$ means to express it
in the form of an ``envelope equation''
\begin{equation}\label{dgl2b}
\ddot{\rho} + \omega^{2}(t)\,\rho -
\frac{g(t)}{\rho^{3}\left(\<x^{2}\>+\<y^{2}\>+\<z^{2}\>\right)} = 0\,,
\end{equation}
with the ``average focusing function'' $\omega^{2}(t)$ defined as
\begin{displaymath}
\omega^{2}(t) = \frac{\omega_{x}^{2}\<x^{2}\>+
\omega_{y}^{2}\<y^{2}\>+\omega_{z}^{2}\<z^{2}\>}
{\<x^{2}\>+\<y^{2}\>+\<z^{2}\>}\,.
\end{displaymath}
Eq.~(\ref{dgl2b}) is equivalent to (\ref{dgl2a})
if the derivative of $g(t)$ satisfies
\begin{align}
\dot{g}(t) = 2\rho^{3}\bigg(&\<xp_{x}\>(\ddot{\rho}+\omega_{x}^{2}\rho) +
\<yp_{y}\>(\ddot{\rho}+\omega_{y}^{2}\rho) +\notag\\
&\<zp_{z}\>(\ddot{\rho}+\omega_{z}^{2}\rho) -
\frac{W}{mN}\dot{\rho}\bigg)\,.\label{dgl2c}
\end{align}
Expressed in terms of $\rho(t)$ and $g(t)$,
the invariant (\ref{invar1a}) writes
\begin{align*}
I_{\rho}/N = & \,\<{\left(\rho\,p_{x}-\dot{\rho}\,x\right)}^{2}\>+
\<{\left(\rho\,p_{y}-\dot{\rho}\,y\right)}^{2}\>\\
&\mbox{} + \<{\left(\rho\,p_{z}-\dot{\rho}\,z\right)}^{2}\>+
\rho^{2}\frac{2W}{mN}+\frac{g(t)}{\rho^{2}}\,.
\end{align*}
Figure~\ref{fig:f2quadr} displays the function $\rho(t)$ resulting from
a numerical integration of the coupled set of differential equations
(\ref{dgl2b}) and (\ref{dgl2c}) for given focusing functions
$\omega_{x}^{2}(t)$, $\omega_{y}^{2}(t)$, $\omega_{z}^{2}(t)$
and initial conditions $g(0)$, $\rho(0)$, $\dot{\rho}(0)$.
The time-dependent coefficients contained herein, namely the second-order
beam moments as well as the field energy $W(t)$ were obtained from a
three-dimensional simulation of a charged particle beam propagating through
a periodic focusing lattice with non-negligible Coulomb interaction,
as described by the potential function (\ref{effpot}).
We observe that for appropriate initial conditions the obtained
evolution of $\rho(t)$ is approximately periodic, as imposed by
the cell length of the periodic focusing lattice.

With regard to Eq.~(\ref{dgl2b}), we may interpret the
function $\rho(t)$ as a ``generalized beam envelope''.
Since the individual inter-particle forces are included in (\ref{effpot}),
non-Liouvillean effects~\cite{struck} emerging from the
granular nature of charge distributions are also covered.

As an outlook, we point out that a major benefit of our result may be derived
in the realm of numerical simulations of systems described by (\ref{ham0}).
Eq.~(\ref{invar1a}) embodies a time integral of Eq.~(\ref{dgl1a}),
provided that the phase-space flow of the particle ensemble is
{\em strictly\/} consistent with the equations of motion (\ref{speqm0}).
This strict consistency can never be accomplished if the time evolution of
the particle ensemble is obtained from a computer simulation
because of the generally limited accuracy of numerical methods.
Under these circumstances, the quantity $I_{f_{2}}$ as given by
Eq.~(\ref{invar1a}) --- with $f_{2}(t)$, $\dot{f}_{2}(t)$,
and $\ddot{f}_{2}(t)$ following from (\ref{dgl1a}) ---
can no longer be expected to be strictly constant.
The deviation of a numerically obtained $I_{f_{2}}$ from a
constant of motion may thus be used as {\em a posteriori\/}
error estimation for the respective simulation.

We finally note that the procedure to derive a quantity $I$ that is conserved
along a system's phase-space path can straightforwardly be
generalized on the basis of (\ref{invar0}) to potentials
with quadratic velocity dependence.

\end{document}